\documentclass[preprint]{vgtc}               

\graphicspath{{figures/}{pictures/}{images/}{./}} 

\usepackage{times}                     

\usepackage{tabu}                      
\usepackage{booktabs}                  
\usepackage{lipsum}                    
\usepackage{mwe}                       

\usepackage{mathptmx}                  

\usepackage{color,soul}

\usepackage[tagged, highstructure]{accessibility}

\usepackage[T1]{fontenc}

\onlineid{1277}

\vgtccategory{Research}

\vgtcinsertpkg

\preprinttext{To appear in IEEE Visualization and Visual Analytics (VIS), 2025.}

\newcommand{\csubref}[2]{%
\crefformat{figure}{##2Fig.~##1.#2##3}
\cref{#1}%
\crefformat{figure}{##2Fig.~##1##3}
}

\NewDocumentCommand{\anon}{m m}
{#1}

\title{Animated Visual Encoding and Layer Blending for \\ Identification of Educational Game Strategies}

\author{
    Braden Roper
    \thanks{e-mail: bradenroper@ou.edu}\\ %
    \parbox{1.4in}{\scriptsize \centering K20 Center \\ The University of Oklahoma }
\and 
    William Thompson
    \thanks{e-mail: will.thompson@ou.edu}\\ %
    \parbox{1.4in}{\scriptsize \centering K20 Center \\ The University of Oklahoma }
\and 
    Chris Weaver\thanks{e-mail: cweaver@ou.edu}\\ %
    \parbox{1.4in}{\scriptsize \centering School of Computer Science \\ The University of Oklahoma }
}

\teaser{
  \centering
  \includegraphics[width=\linewidth]{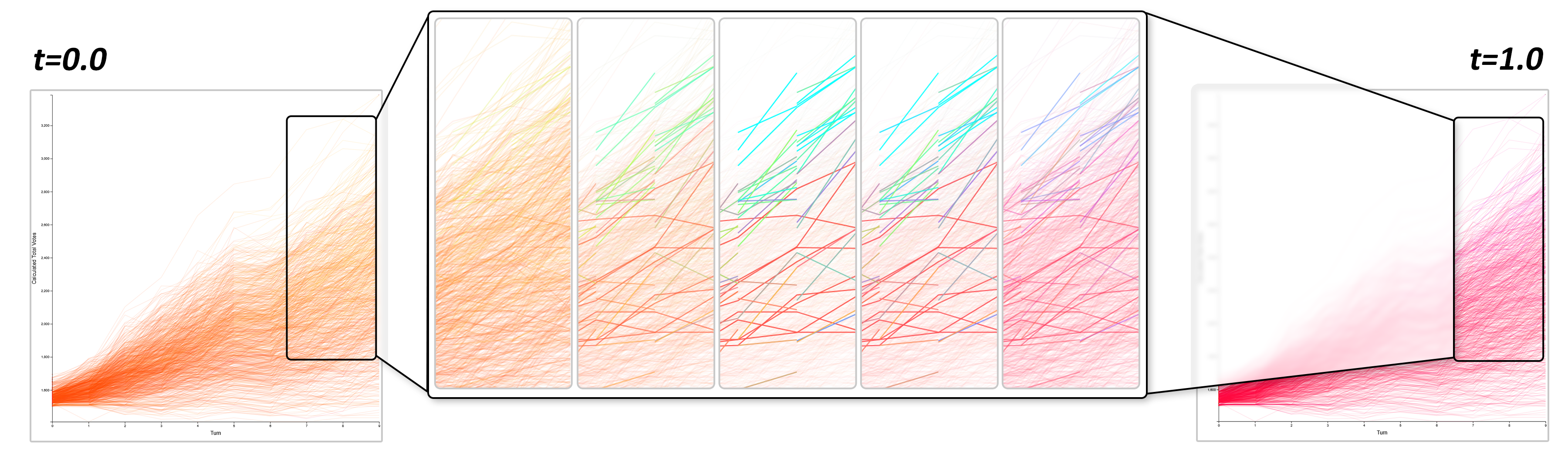}
  \caption{
  An example of animated output of a kinetic query, revealing a successful strategy in the election-simulation and computational thinking educational game \anon{Operation: ELECT.}{Anonymous Game.} The line plot shows calculated acquired votes over the turns of an individual level. Isolated success in two key districts of the specific in-game level are encoded with red-to-green and red-to-blue color scales configured to overlap near the midpoint of the animation, creating a cyan blended encoding for players who were successful in both districts. An opacity encoding is also blended in near the midpoint of the animation to accentuate in-game actions of interest.
  }
  \alt{Output from the visualization tool is shown as a color and opacity encoded line plot. Two whole plots are shown on the left and right at the beginning and end of the animation with a callout showing multiple frames in between, in an attempt to display the technique's animated nature.}
  \label{fig:teaser}
}

\abstract{

Game-Based Learning has proven to be an effective method for enhancing engagement with educational material.
However, gaining a deeper understanding of player strategies remains challenging.
Sequential game-state and action-based tracking tools often gather extensive data that can be difficult to interpret as long-term strategy.
This data presents unique problems to visualization, as it can be fairly natural, noisy data but is constrained within synthetic, controlled environments, leading to issues such as overplotting which can make interpretation complicated.
We propose an animated visual encoding tool that utilizes kinetic visualization to address these issues.
This tool enables researchers to construct animated data narratives through the configuration of parameter interpolation curves and blending layers.
Finally, we demonstrate the usefulness of the tool while addressing specific interests as outlined by a domain expert collaborator.

} 


\begin{document}




\firstsection{Introduction}
\maketitle


In recent years, a need for computational thinking and data literacy education has been identified, with more jobs requiring familiarity with data and advanced software systems, even outside the field of Computer Science~\cite{campGenerationCSGrowth2017}.
Educational standards~\cite{CSTA2017} and models~\cite{hunsaker2023ctmodel} are new and still being refined.
The \anon{K20 Center's}{Anonymous Institution's} educational game \anon{Operation: ELECT~\cite{operationElectGame}}{Anonymous Game} was developed to aid in this new computational thinking education, while also connecting to social studies standards for electoral processes in \anon{the United States}{Anonymous Country}.
In the game, students play as campaign managers who work to get their candidate elected.
They are presented with simulated polling data, news events, and regional interests that they interpret to inform their decisions.
To be successful, a player cannot simply address the specific circumstances of a given turn, but must also develop a long-term strategy.
The game's primary Instructional Designer---and second author of this paper---is especially interested in analyzing and visualizing player strategy to evaluate the balance of game mechanics and to confirm the game's efficacy as an educational resource.

In this paper, we explore the use of animated visualization as a means to reveal patterns in complex gameplay data.
If a simple line chart can represent per-player progression of a single state variable, then it follows that additional visual encoding can be applied to per-turn line segments to depict other analytically relevant state variables.
The main building blocks of all graphical elements in visualization systems, known as visual encodings, typically include attributes such as position, size, shape, color, and orientation.
For this application we focus solely on color and opacity due to their flexibility and blending properties when encoding information onto an underlying line plot.
The proposed technique extends these channels through animation to offer a new method to analyze complex game data.
This is done through the introduction of \emph{kinetic queries}, which can be thought of as an extension of dynamic queries~\cite{ahlberg1994visual, ahlberg1992dynamic}.
In addition to specifying the data parameters and the encoding channels that will be used, kinetic queries include an animation curve by which the encoding will be applied and the blending mode used to accumulate the overall visual mapping.
This gives analysts the flexibility to stack or sequence visual effects.

To demonstrate the usefulness of the proposed tool, we perform example analyses on gameplay data.
Key interests are identified by our domain expert collaborator and aligned to the three provided examples.
Although the given analyses focus solely on encoding information into the color and opacity of animated line charts, our ultimate goal is to further develop this technique and include other encoding channels and types of visualizations.

\begin{figure*}[t!] 
    \centering
    \includegraphics[width=\textwidth]{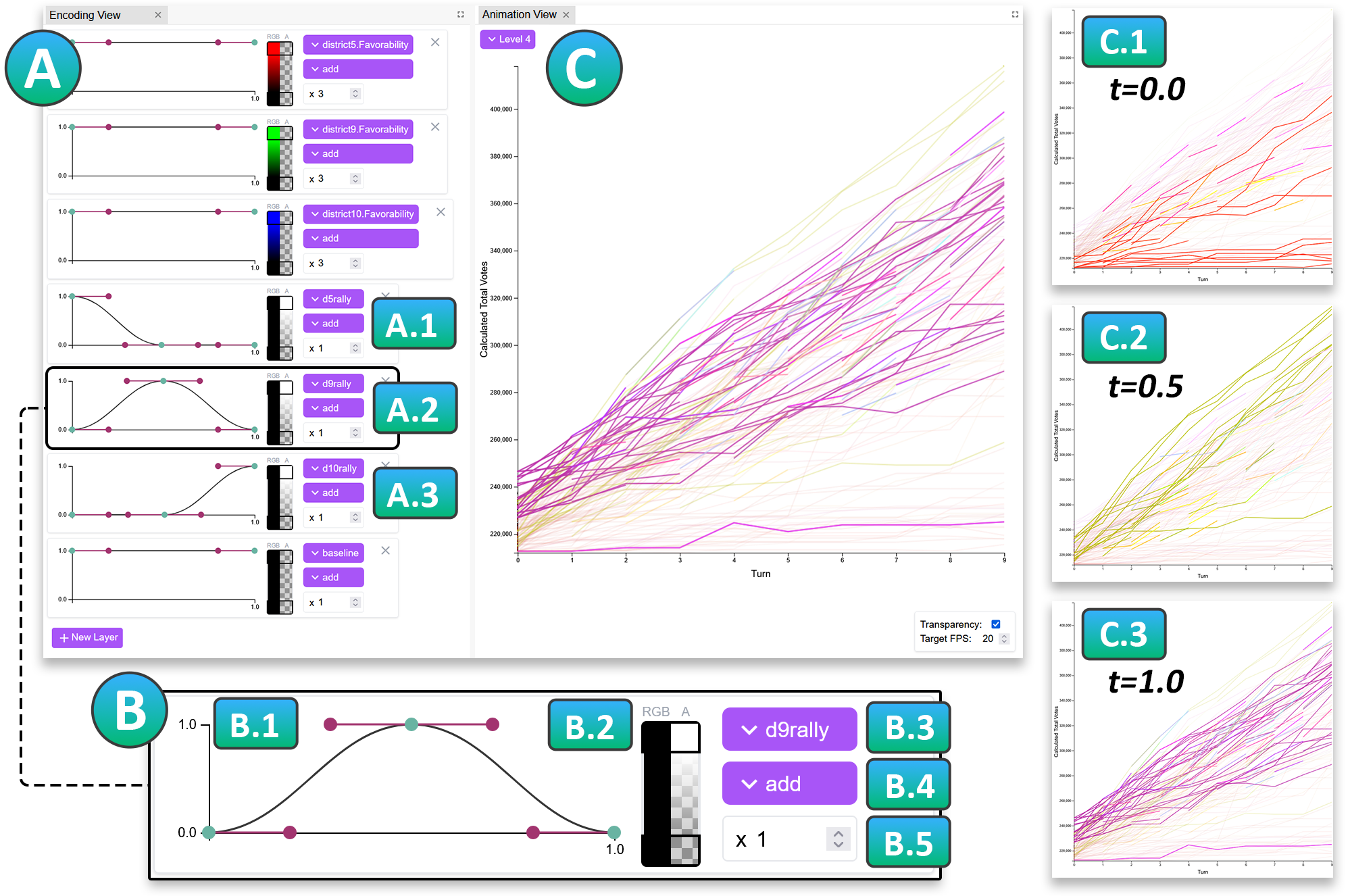}
    \caption{
    The full interface of the tool has two main views: the Encoding View and the Animation View. The Encoding View (A) allows analysts to add and configure layers, one of which (B) is expanded here to show its inputs (B.1-5) in greater detail. The Animation View (C) continuously plays an animated visual based on the configured layers. The full effect of the technique cannot be seen here due to its animated nature. 
    However, multiple time-steps (C.1-3) can be seen at times $t=0.0$, $t=0.5$ and $t=1.0$, where the animation curves in A.1-3 are at their peaks.
    }
    \alt{The full interface of the tool is shown, along with a zoomed in configuration for a single blending layer and snapshots of the output at three different time steps. The full interface is made up of a sequence of user-configured blending layers on the left and the outputted line graph on the right.}
    \label{fig:interface}
\end{figure*}

\section{Related Work}

Game-Based Learning activities have been prized for their ability to increase engagement, adapt to the circumstances of a specific player, and allow graceful failure~\cite{hilton2011learning,plassFoundationsGameBasedLearning2015}.
Successful implementation of these games requires careful instructional design, which often includes proper incorporation of educational material, a balance of player cognitive load, and the right amount of guidance~\cite{tobias2007research,tobiasGameBasedLearning2014}.
Much of these design skills and templates are powered by analyzing game data that, in recent years, has greatly increased in volume and complexity, prompting research in the evolving field of Learning Analytics~\cite{freireGameLearningAnalytics2016, romero2020educational}.



Although there are many examples of visualization for game data~\cite{medlerAnalyticsPlayUsing2011,mouraVisualizingUnderstandingPlayers2011,ruiperez-valienteIdeatingDevelopingVisualization2021,wallner14IntroductionGameplay}, representing complex game states and their changes over time remains difficult.
Such difficulties have led us to explore animation techniques such as kinetic visualization, designed to utilize our perceptual abilities to detect patterns in motion~\cite{lumKineticVisualizationTechnique2002,lumUsingMotionIllustrate2003}.

Early implementations of kinetic visualization showed promise, such as the use of \emph{moxel} displays to explore geospatial data~\cite{bobrow2005kinetic,yangDataExplorationCombining2006}.
However, further research in the field is surprisingly sparse, and work that does exist often limits the use of the time dimension for the exclusive representation of the temporal aspect of a dataset~\cite{Gapminder, tverskyAnimationCanIt2002a}.
While this use of time is conceptually natural, we wish to look past this restriction and encourage additional exploration into more flexible uses of animation.


\section{Approach}

The proposed technique aims to create animated visualizations by adding kinetic queries in the form of layers.
The tool is divided into two main interfaces: the Encoding View (\csubref{fig:interface}{A}) and the Animation View (\csubref{fig:interface}{C}).

\subsection{Encoding Layers}\label{sec:approach-encoding}
The Encoding View (\csubref{fig:interface}{A}) allows analysts to add any number of layers, each with a dedicated purpose to link a data parameter to a visual effect.
Each layer has a variety of inputs, including an animation curve (\csubref{fig:interface}{B.1}), a color scale (\csubref{fig:interface}{B.2}), a parameter selection dropdown (\csubref{fig:interface}{B.3)}, a blending mode selection dropdown (\csubref{fig:interface}{B.4}), and an additional multiplier (\csubref{fig:interface}{B.5}).
The color scale previews its color channels and transparency channel separately since an analyst may wish to control these independently, and low transparency values by their nature would make a color selection difficult to interpret.
The functional purposes of these inputs, as well as the methods used to combine them, are described in more detail in \cref{sec:approach-pipeline}.

Each layer will output a color object that contains three color channels and a transparency, or alpha, channel.
At a given time step, every layer is processed at the same time $t$ with layers being processed sequentially starting from the top and passing their outputs to the layer below them.
The final layer dictates the visual encoding to apply to a data point in the Animation View (\csubref{fig:interface}{C}).
This means that by scanning vertically across the layers, an analyst can stack the desired visual effects by modifying each layer's animation curve, allowing their kinetic queries to be applied sequentially, in sync, or with any desired amount of overlap.

\subsection{Layer Processing and Blending}\label{sec:approach-pipeline}
\begin{figure}
    \centering
    \includegraphics[width=\linewidth]{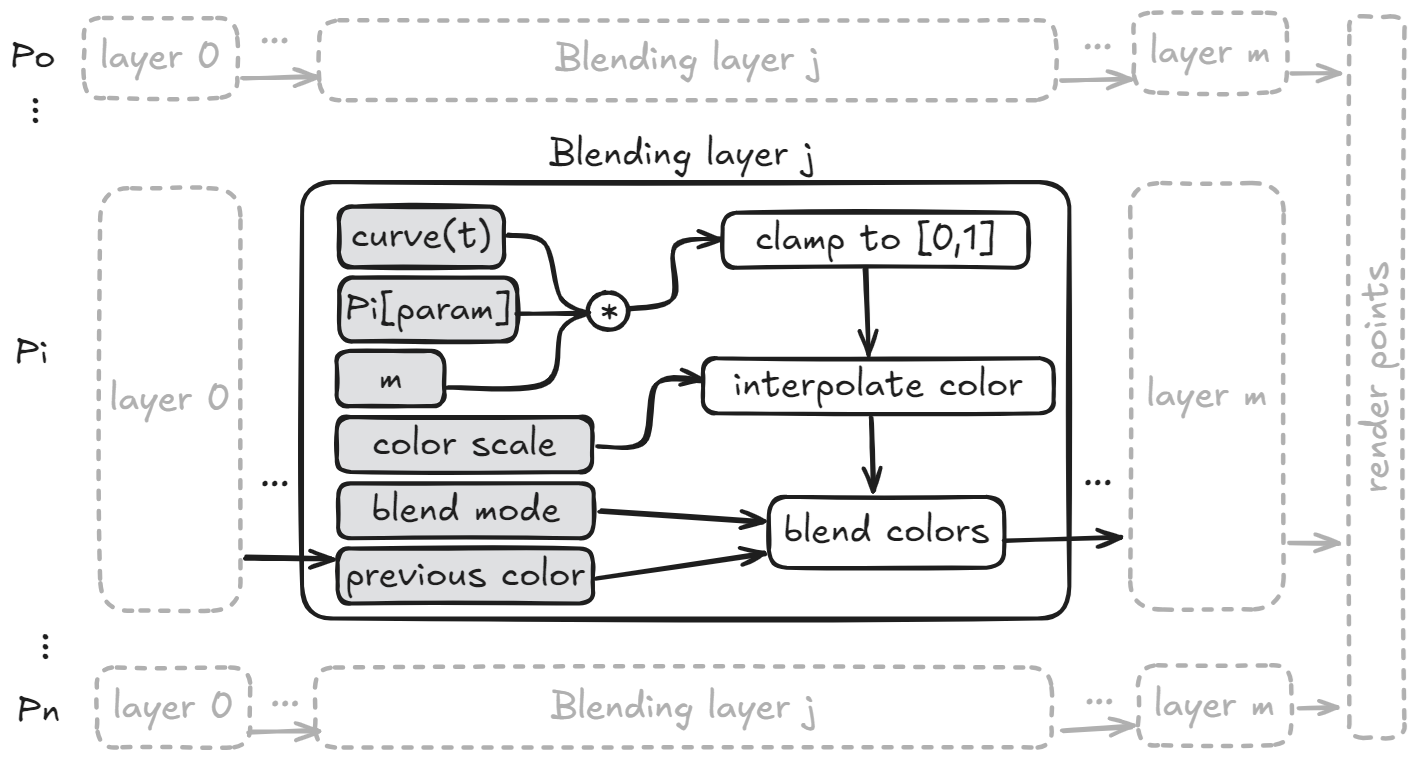}
    \caption{A layer processing diagram is shown. At any time $t$, each point $P_i$ has a calculated visual encoding determined by its sequential flow through every blending layer. The input elements are shaded, with the first five being set through the blending layer's interface. The final input is provided as output from the previous layer. }
    \alt{A processing diagram is shown, outlining the steps each blending layer goes through in order to contribute to the final output. Inputs to each layer are shaded to distinguish them from the nodes representing processing steps.}
    \label{fig:diagram}
\end{figure}

At each time step, each visualized data point will be processed through all defined encoding layers.
Each layer has the inputs outlined in \cref{sec:approach-encoding}, with an additional input color provided as output from the previous layer, as seen in \cref{fig:diagram}.
In the case of the first layer, a constant color is assumed for the previous color, with $0$ values for all color and transparency channels.

To process a data point in a single layer, we must first obtain the interpolation parameter $c$, found by the following equation:

$$
c = A(t)*p*m
$$

where $A(t)$ is the animation curve's value at time $t$, $p$ is the value of the chosen parameter for the current data point (or alternatively a constant \emph{baseline} option), and $m$ is the additional multiplier, which is $1$ by default.
This product $c$ is then clamped to the range of $[0,1]$ and used as the interpolation parameter along the specified color scale (\csubref{fig:interface}{B.2}), resulting in a color value, including transparency.
The specified blending function (\csubref{fig:interface}{B.4}) then combines this color with the result from the previous layer, returning a new color that is passed along to the next layer in sequence or, in the case of the final layer, to the Animation View for rendering.

The blending functions used in this work are inspired by those of image and video editing software~\cite{valentine2012hidden}.
We include three basic blending functions.
\emph{Add} and \emph{Multiply} apply simple arithmetic operators at the level of individual color and alpha channels.
\emph{Mask} takes the minimum alpha value from the two inputs and ignores the main color channels, copying them from the previous layer.



\subsection{Animated Output}\label{sec:approach-animation}
The Animation View (\csubref{fig:interface}{C}) constantly loops over a visualization based on the layers specified in the Encoding View (\csubref{fig:interface}{A}).
For the scope of our provided examples and analyses in \cref{sec:examples}, we use kinetic encoding to enhance line graphs.
Adding new encoding layers, or kinetic queries, will automatically update the visualization, providing quick feedback to analysts.

The additional frames seen in \csubref{fig:interface}{C.1-3} are not part of the interface but are simply the rendered output at different time steps in which the animation curves of a few layers (\csubref{fig:interface}{A.1-3}) are at their peaks, meaning their contributed visual effects are at an extrema.
To observe the animated output of this technique, refer to the provided supplemental video.

\section{Analysis and Findings}\label{sec:examples}

We now demonstrate the use of this technique to explore collected game data from 378 students across Oklahoma.
These students attend schools that are partnered with the \anon{K20 Center}{Anonymous Institution} through \anon{federal GEAR UP grants}{Anonymous Grants}.
\anon{Operation: ELECT}{Anonymous Game} uses a tracking system based on the xAPI standard~\cite{xapiStandard}, which records user actions, the context in which those actions were taken, and the resulting changes to a player's game state. 

For this analysis, we focus on one game level at a time, plotting a player's total votes against their turn and exposing per-turn parameters such as \emph{budget} and \emph{duration}, as well as a set of parameters for each district.
On each turn, every district has: a \emph{population}, a candidate \emph{favorability} score,  percentages of population that are \emph{unregistered} or \emph{undecided}, percentages of population that are \emph{for} or \emph{against} the player's candidate, and finally a binary label for each type of \emph{action}, with $1$ representing that the action was taken in the given district on the given turn and $0$ representing that it was not.

\subsection{Key Interests}

The primary Instructional Designer of \anon{Operation: ELECT}{Anonymous Game}---and second author of this paper---identified a few Key Interests (KIs) when considering strategic analysis of players, efficacy of the game as an educational activity, and balance of the game's mechanics.

\textbf{KI.1}
While assumptions can be made about the use of displayed district data in player decision-making, it is difficult to interpret long-term strategy spanning multiple turns of the game.
Identification of strategy is especially crucial for \anon{Operation: ELECT}{Anonymous Game}, as it corresponds to the algorithmic thinking that is core to educational computational thinking standards.

\textbf{KI.2}
We are interested in comparing the use of individual actions and highlighting any common combinations of actions.
It is also of interest to identify any correlation between individual or combinations of actions and when they are taken in a playthrough.

\textbf{KI.3}
In general, we would like to identify any areas of imbalance in the game.
While it is intended for players to identify useful and successful strategies, imbalanced game mechanics may hinder learning objectives.

\textbf{KI.4}
We are interested in districts that have high, or highly patterned, activity.
For instance, frequent action-district pairs may indicate common solutions or mistakes that the designer would like to be aware of.

\textbf{KI.5}
As is common in educational game research, we are interested in the efficacy of the game as a learning tool.
For \anon{Operation: ELECT}{Anonymous Game}, one indication of this efficacy is the convergence of strategic behaviors in players, observed as a gradual reduction in exploratory behavior and an increase in exploitative strategies.

\subsection{Examples}
Based on the Key Interests outlined above, we analyzed three examples using the proposed technique.

\textbf{Example 1.} 
Using the tool, successful playthroughs of level one reveal significant activity in districts one and two (\textbf{KI.4}).
This makes sense as these districts initially contain higher percentage of population against the player's candidate, meaning player focus in these districts is more fruitful when attempting to achieve the popular vote.
In addition, many successful players used fundraiser actions in district four, where players are likely to get more money due to their initially high support.

As seen in \cref{fig:teaser}, the tool also aids in visualizing these phenomena by encoding red-to-green and red-to-blue color scales on the percentages of population \emph{for} the player's candidate in these districts.
High values from both color scales (green and blue) form the resulting cyan region near the middle of the animation.
Masking layers are also added to encode the use of specific actions, utilizing opacity to highlight a particularly successful combination of actions~(\textbf{KI.2}).
Namely, we encode high opacity on turns where users exercise grassroots actions in district one and two and a fundraiser action in district four. The resulting visualization reveals a strategy that shows success in districts one and two, which unsurprisingly correlates with higher total votes for the level.


\textbf{Example 2.}
In level four, we first noticed two visible clusters of lines.
Most move upwards as players gain votes while some remain largely horizontal (see \csubref{fig:interface}{C}, with the bottom cluster most visible in \csubref{fig:interface}{C.1}).
We quickly discovered shorter turn durations for this lower cluster, possibly identifying students who hurried through the level and, as a result, made less informed decisions.

Continuing our analysis, we focus on rally actions in districts five, nine, and ten (\textbf{KI.2}).
Our Instructional Design expert showed special interest in the use of rallies because they are crucial for increasing \emph{favorability} and are expensive, requiring an \emph{appearance} (another type of in-game currency) in addition to their monetary cost.
As explained to players in the game: higher \emph{favorability} scores result in more effective actions and cause more \emph{undecided} voters to swing your way on election day.

We encode red, green, and blue color channels for \emph{favorability} in these districts (see first three layers in \csubref{fig:interface}{A}).
We give them flat animation curves so their colors remain constant in our animation, unlike the layers seen in \csubref{fig:interface}{A.1-3}, whose curves animate opacity where rally actions are taken.
The resulting animation contains three main phases: (\csubref{fig:interface}{C.1}) district five rallies are shown, with a mostly red-orange hue, (\csubref{fig:interface}{C.2}) district nine rallies are shown, with a mostly yellow hue, and (\csubref{fig:interface}{C.3}) district ten rallies are shown, with a mostly magenta hue.
Note that since colors are actually static in this animation, their apparent change is caused by the fading in and out of data points based on the rally actions in different districts.
Interestingly, in addition to their own district, the latter two phases show that players have high \emph{favorability} in district five (seen as yellow and magenta, which are formed with a significant red channel contribution), as opposed to the first phase which shows a high \emph{favorability} in district five but not in the other two districts.
Additionally, these two phases show generally higher total votes, meaning these players' strategies were more successful~(\textbf{KI.1}).

From this visualization, a few conclusions can be drawn.
First, spending valuable resources on rallies in district five is less successful than using the same action in districts nine and ten.
More interestingly, there is very little overlap between the successful players who focused on districts nine and ten, which can be seen in the later half of the animation as cyan lines (with high green and blue channels).
This is promising for the efficacy of the game as a computational thinking learning tool, as the players found success in various ways by committing to different long term strategies~(\textbf{KI.5}).


\textbf{Example 3.} 
Our expert is also interested in analysis of game balance.
Using the tool, we can explore individual actions, overlap them with other actions, and localize them to specific districts.
Through such exploration, we identified an unexpectedly frequent use of the \emph{voter registration drive} action among higher-scoring players (\textbf{KI.3}), especially in the last three levels of the game.
We used additive layers encoding color and full opacity for the voter drive action in every district, along with a baseline layer adding a small amount of opacity to all lines, producing an animated pulse that highlights turns with these actions.
In an attempt to isolate high voter drive usage in specific districts, we divided districts into groups of two to four (depending on the total number of districts in the level we were analyzing) and assigned groups dedicated color channels, with the plan to drill down into the groups of districts with the higher represented colors.
To our surprise, even after shuffling color usage and groups, the animations always remained very noisy and colorful, due to this action being taken in most districts, and often across two districts in one turn.

The wide use of this action is not necessarily an issue, but it was designed to be more situational, and therefore deserves some analytical attention.
It is also possible that this action is used by players who have already been successful and are simply using up their extra budget.
Although it does not appear to cause steep increases in overall votes when used, further analysis may be needed to determine if it is due to imbalance or simply a tool that users have successfully incorporated into their strategies.


\begin{figure}
    \centering
    \includegraphics[width=\linewidth]{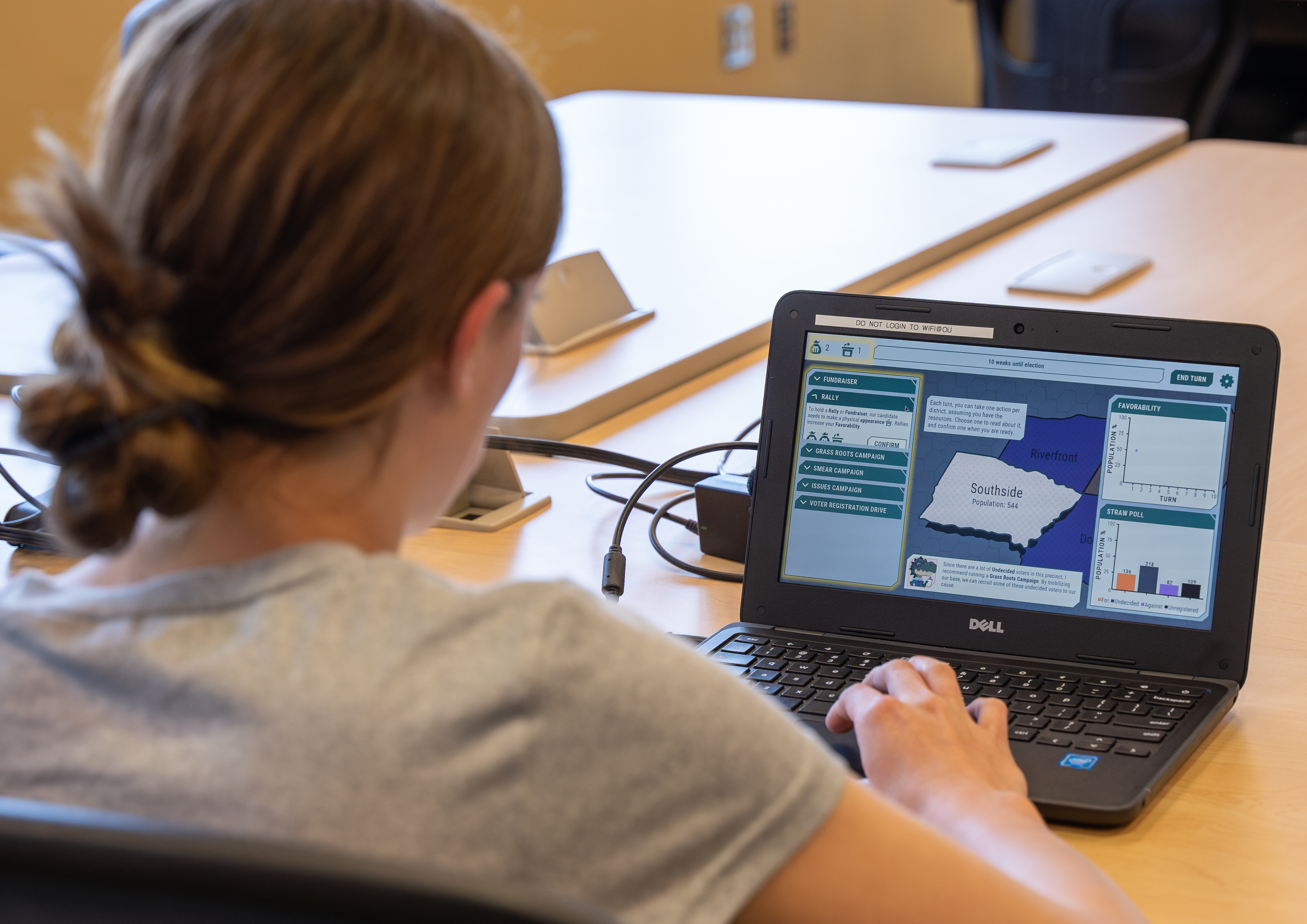}
    \caption{
    A \anon{GEAR UP}{Anonymous Grant} cohort student plays \anon{Operation: ELECT}{Anonymous Game}, viewing one of four districts in level one. The left side of the screen shows available actions that may be taken in the selected district each turn. The right side of the screen shows district analytics the player must interpret to make informed decisions. 
    }
    \alt{A student from the K20 Center's GEAR UP for LIFE grant cohort is shown playing the educational game, Operation: ELECT.}
    \label{fig:student-usage}
\end{figure}

\section{Conclusions and Future Work}

We present a kinetic query visualization design and its useful application in the analysis of educational game data.
Animated visualizations generated with the implemented tool demonstrate its exploratory capabilities and utility for visual analysis.
Specification of kinetic queries in the tool is relatively simple and straightforward, but also suggests conveniences and extensions to enhance the effectiveness of the animated encodings.
We hope to achieve greater flexibility by incorporating logical grouping of layers, fine-grain control over other visual encoding channels, and more robust temporal configurations.
We also plan to include more operators and blending functions to allow for more animation effects and address issues such as overplotting.
More functional operators could address one particular interest, as stated by our domain expert, to highlight phenomena in later turns based on specific conditions of earlier turns.
These additions would make the technique more robust, supporting the analysis of other educational games and other domains in general.

The kinetic queries technique shows promise in revealing patterns in complicated data.
It may also serve as a basis for the development of presentation and data storytelling techniques. While animation performance was sufficient in our game data application, it would likely be an issue for more complex query expressions and larger amounts of data. 
Moving forward, we plan to study the limitations of the technique's effectiveness and performance as we expand the expressiveness of its kinetic querying features beyond the color channels and blending modes that we used to animate the line chart in this particular application.


\acknowledgments{
The authors wish to thank the \anon{K20 Center at the University of Oklahoma}{Anonymous Institution}, whose Game-Based Learning activity and data supported this research.
This work was supported in part by the \anon{K20 Center's}{Anonymous Institution's} \anon{GEAR UP for LIFE grant, GEAR UP for OKC grant, GEAR UP for the FUTURE grant, and GEAR UP for MY SUCCESS grant}{Anonymous Grant}.
}

\bibliographystyle{abbrv-doi}

\bibliography{template}
\end{document}